# High efficiency spin-decoupled modulation using chiral C2-symmetric meta-atoms

**Author:** Haohan Chen,[1] Jiepeng Wu,[1] Minglei He,[1] Hao Wang,[1] Xinen Wu,[1] Kezhou Fan,[2] Haiying Liu,[1] Qiang Li,[1] Lijun Wu,[1*] and Kam Sing Wong[2]

[1]*Guangdong Provincial Key Laboratory of Nanophotonic Functional Materials and Devices, School of Information and Optoelectronic Science and Engineering, South China Normal University, Guangzhou 510006, China*
[2]*Department of Physics and William Mong Institute of Nano Science and Technology, The Hong Kong University of Science and Technology, Hong Kong 999077, China*
*Corresponding author: ljwu@scnu.edu.cn

**Abstract:** Orthogonal circularly polarized light is essential for multiplexing tunable metasurfaces. Mainstream spin-decoupled metasurfaces, consisting of numerous meta-atoms with mirror symmetry, rely on the cooperative modulation of the Pancharatnam-Berry (PB) phase and the propagation phase. This paper demonstrates spin-decoupled functionality through the synergistic utilization of planar chiral meta-atom phase response and PB phase. Based on the Jones calculus, it has been found that meta-atoms with chiral C2-symmetry owns a larger geometric parameter range with high cross-polarization ratio compared to those with mirror symmetry or higher symmetries at the same aspect ratio. This characteristic is advantageous in terms of enabling high-efficiency manipulation and enhancing the signal-to-noise ratio. As an example, 10 kinds of C2-symmetry chiral meta-atoms with a H-like shape are selected by the self-adaptive genetic algorithm to attain a full $2\pi$ phase span with an interval of $\pi/5$. To mitigate the additional propagation phase change of the guided modes originated from the arrangement alternation upon the rotation of the meta-atoms, the enantiomer of chiral meta-atoms and its PB phase delay are adopted to minimize the difference between the actual and desired target phases. A polarization-insensitive metalens and a chiral virtual-moving metalens array are designed to demonstrate the spin-decoupled function with both high efficiency and signal-to-noise ratio. The work in this paper may trigger more exciting and interesting spin-decoupled multiplexing metasurfaces and broaden the prospect of chiroptical applications.

# 1. Introduction

Metasurfaces [1], consisting of planar periodic subwavelength artificial meta-atoms, have attracted unprecedented attention because of the flexibility to manipulate light. Unlike conventional optical devices that rely on phase accumulation, metasurfaces provide a novel platform for electromagnetic wave manipulation through the generalized Snell's law [2,3]. Dozens of fascinating functionalities have been proposed, such as metalens [4,5], holograms [6,7], vortex beam generators [8,9], to name but a few. In recent years, versatile multiplexing tunable metasurfaces have been demonstrated by active [10-13] or passive [6,7,14-18] switching. Comparing with active switching, passive switching metasurfaces are easier to fabricate and exhibit more stable performances. Among the various passive switching methods, polarization serves as a key parameter for convenient and robust modulation through the use of a polarizer.

Typically, polarization-based multiplexing metasurfaces, comprised of anisotropic nano-antennas, can be independently modulated through orthogonal polarization channels (e.g., orthogonal linear [14], circular [6,7] or elliptical [7] polarization). Among these channels, the independent modulation of orthogonal circular polarization holds great importance in various applications, including dual-channel hologram [6,7], pancake metalens [19], vector visual cryptography [20] and more. Considerable efforts have been demonstrated to realize the spin-decoupled function, employing various approaches such as the combination of propagation phase and Pancharatnam-Berry (PB) phase [7,20], area-division based on PB phase [21,22], planar chiral meta-atoms [6,23], the combination of Aharonov-Anandan phase and PB phase [24], and additional methods [25,26]. For the aforementioned modulation methods, the presence of the non-modulated polarization component noticeably contributes to the main background noise, thereby deteriorates the signal-to-noise ratio.

Increasing the number of meta-atoms with a high cross-polarization ratio (CPR) [21] is crucial for improving the phase modulation efficiency and signal-to-noise ratio of metasurfaces. For dielectric metasurfaces, it is evident that utilizing materials with high refractive index or increasing the aspect ratio of meta-atoms can expand the pool of meta-atoms with both high transmission and high CPR [4,27]. Nevertheless, the fabrication of meta-atoms with large aspect ratios poses a substantial challenge. Another method involves broadening the diversity of the basic shape of meta-atoms in searching for more appropriate shapes [27,28], which in turn increases the computational resource requirements. Consequently, there is a significant demand for identifying design restrictions on the basic shape to increase the quantity of high-efficiency meta-atoms at a specific aspect ratio, which can shorten the optimization process. On the other hand, structural symmetry plays a fundamental role in determining the physical properties of meta-atoms [29]. For example, various symmetrical meta-atoms result in different PB phase delays, both in linear [30] and nonlinear [31] regions. The symmetry of meta-atoms could be a vital parameter that impacts the abundance of meta-atoms with a high CPR.

In this work, based on the Jones calculation, we explore the influence of the structural symmetry on the geometric parameter range of the meta-atoms with high CPR. For a given basic shape, our findings indicate that planar meta-atoms with chiral C1- and C2-symmetries exhibit a greater quantity of instances with high CPR compared to those with mirror and higher symmetries, whereby phase modulation efficiency and signal-to-noise ratio can be increased. Compared to C1-symmetry, C2-symmetry requires fewer computing resources due to the reduced number of variables associated with a fixed basic shape. As an example, this paper utilizes a self-adaptive genetic algorithm to carefully select 10 types C2-symmetry chiral meta-atoms with a H-like shape. By employing PB phase modulation and optimizing through its enantiomer, these 10 kinds of chiral meta-atoms enable the effective implementation of the spin-decoupled modulation function by covering the entire $2\pi$ phase range with an interval of $\pi/5$ for the orthogonal circular polarization (CP). The design strategy employed in this spin-decoupled phase library effectively reduces the number of required meta-atoms, thereby minimizing the design complexity and reducing the processing time. As any polarized light can be decomposed into two orthogonal CPs, a polarization-insensitive metalens is designed using the phase library, with both types of CP maintaining the same phase response. In addition, by manipulating the phase delay of these orthogonal CPs, a chiral virtual-moving metalens array is constructed, which can improve the spatial resolution of light-field imaging [14,32]. Our work will enrich the domain of high efficiency spin-decoupled modulation metasurface and may prompt more wonderful chiroptical applications.

## 2. CPR of different symmetric meta-atoms

As illustrated in Fig. S1(a), anisotropic nano-antennas with mirror symmetry (e.g., nanorod) exhibit a limitation wherein only the induced current flowing in the same direction can be excited when a linearly polarized electric field impinges along one of their optical axes. As a result, they are unable to generate disparate phase shifts for orthogonal CP light. In contrast, in nano-antennas with broken mirror symmetry, as depicted in Fig. S1(b), the induced current flow in the orthogonal direction can be excited. This capability potentially results in distinct phase shifts for different CP light.

In the case of subwavelength anisotropic nano-antennas, the excited electric field is commonly modeled as an electric dipole, where the dipole moment is assumed to be parallel in the *x-y* plane. When CP light is incident, the perpendicular radiation electric field $E_t$ can be derived as (See Sec. S1 of Supplemental Material for more details [33]):

$$\vec{E}_t = \begin{cases} A_0\left(p_{LL}\vec{s}_L + p_{RL}\vec{s}_R\right) = A_0\left(A_1 e^{j\varphi_1}\vec{s}_L + A_2 e^{-j\varphi_2}\vec{s}_R\right), & I_L \\ A_0\left(p_{LR}\vec{s}_L + p_{RR}\vec{s}_R\right) = A_0\left(A_2 e^{j\varphi_2}\vec{s}_L + A_1 e^{-j\varphi_1}\vec{s}_R\right), & I_R \end{cases} \quad (1)$$

$$\begin{cases} A_0 = \dfrac{-\omega^2}{4\pi\varepsilon_0 c^2 R_0} e^{jkR_0} \\ A_1 = \dfrac{\varepsilon_0}{2}\sqrt{\left(\chi_{xx}+\chi_{yy}\right)^2 + \left(\chi_{xy}-\chi_{yx}\right)^2} \\ A_2 = \dfrac{\varepsilon_0}{2}\sqrt{\left(\chi_{xx}-\chi_{yy}\right)^2 + \left(\chi_{xy}+\chi_{yx}\right)^2} \\ \varphi_1 = \arctan\dfrac{\chi_{xy}-\chi_{yx}}{\chi_{xx}+\chi_{yy}} \\ \varphi_2 = -\arctan\dfrac{\chi_{xy}+\chi_{yx}}{\chi_{xx}-\chi_{yy}} \end{cases}, \quad (2)$$

where $p_{LL}$, $p_{LR}$, $p_{RL}$ and $p_{RR}$ are the electric dipole moments in circular base. The subscript of *R/L* refers to RCP/LCP. $I_L$ and $I_R$ represent RCP and LCP incidences. $\chi_{xx}$, $\chi_{xy}$, $\chi_{yx}$ and $\chi_{yy}$ are the electric polarizability tensors in linear orthogonal base. $\vec{s}_L$ and $\vec{s}_R$ are the unit vectors in circular base of LCP and RCP, respectively. $R_0$ is the distance between the observation point and the electric dipole. $c$ and $\varepsilon_0$ are the velocity of light and the permittivity in vacuum.

Eq. (1) shows that the amplitude of the transmitted electric field, which exhibits the same/opposite chirality to the incident light, is given by $A_0A_1/A_0A_2$. For a planar chiral meta-atom without mirror symmetry, the cross-component polarizability tensor is $\chi_{xy}\neq 0$ and $\chi_{yx}\neq 0$. Under LCP(RCP) illumination, the transmission of light with the same/opposite handedness compensates the phase of $\varphi_1(-\varphi_1)$ / $-\varphi_2(\varphi_2)$, showcasing the spin-decoupled function of planar chiral meta-atoms in a manner reminiscent of PB phase modulation.

CPR is defined as [21]:

$$CPR_{RL(LR)} = \frac{T_{RL(LR)}}{T_{RL(LR)} + T_{LL(RR)}}, \quad (3)$$

where $T_{RL}$, $T_{LR}$, $T_{RR}$ and $T_{LL}$ represent the transmittance for different combinations of incident (the second subscript letter) and transmitted (the first subscript letter) polarizations. For example, $T_{RL}$ represents the transmittance of RCP for LCP incident. All the same subscripts in the whole paper represent the same meaning unless otherwise specified.

Combining Eq. (1) and (2), CPR can be deduced to be:

$$CPR = \frac{A_2^2}{A_1^2 + A_2^2} = \frac{\left(\chi_{xx} - \chi_{yy}\right)^2 + \left(\chi_{xy} + \chi_{yx}\right)^2}{\left(\chi_{xx} + \chi_{yy}\right)^2 + \left(\chi_{xy} - \chi_{yx}\right)^2 + \left(\chi_{xx} - \chi_{yy}\right)^2 + \left(\chi_{xy} + \chi_{yx}\right)^2}. \quad (4)$$

For mirror symmetric meta-atom, the electric polarizability tensor satisfies $\chi_{xy} = \chi_{yx} = 0$, $\chi_{xx} \neq \chi_{yy} \neq 0$; C1-symmetric meta-atom without mirror symmetry satisfies $\chi_{xx} \neq \chi_{yy} \neq \chi_{xy} \neq \chi_{yx} \neq 0$; C2-symmetric meta-atom without mirror symmetry satisfies $\chi_{xy} = \chi_{yx} \neq 0$, $\chi_{xx} \neq \chi_{yy} \neq 0$; C$n$-symmetric ($n$>2) meta-atom satisfies $\chi_{xy} = -\chi_{yx}$, $\chi_{xx} = \chi_{yy}$ [29]. Based on Eq. (4), the value of CPR with different symmetries can be described as follows (See Sec. S2 of Supplemental Material for more details [33]):

$$CPR_{C_1}, CPR_{C_2} > CPR_M > CPR_{C_n}, \quad (n > 2). \quad (5)$$

The polarization conversion efficiency (PCE) of meta-atoms can be obtained by calculating the product of CPR and the transmittance, with the latter being primarily determined by material properties. Maximizing CPR is therefore crucial to improve PCE of the structure. Indeed, it is essential to acknowledge that Eq. (5) is valid only under the assumption that $\chi_{xx} \neq \chi_{yy} \neq \chi_{xy} \neq \chi_{yx} \neq 0$ is satisfied and the polarization tensors of each symmetric structure are identical. However, satisfying the latter assumption is often challenging. In other words, Eq. (5) implies that within a given basic shape and parameter space, the number of meta-atoms with chiral C1- and C2-symmetry exhibiting high CPR is larger compared to those with mirror or other higher symmetric counterparts. As an example, Fig. 1 illustrates the CPR of various meta-atoms with different symmetries. Further information regarding the PCE and transmittance of these meta-atoms can be found in Figs. S2 and S3, respectively. For the sake of a straightforward comparison, the parameters have been simplified to $L$ and $W$ ($L>W$). As shown, the transmittance of various symmetric meta-atoms shows similarities (Fig. S3). However, it is obvious that C1 and C2-symmetric chiral meta-atoms have a notably larger region with high CPR compared to meta-atoms with mirror and higher symmetries (Fig. 1). Consequently, C1 and C2-symmetric chiral meta-atoms demonstrate significantly higher PCE within this extended region compared to those with mirror and higher symmetries (Fig. S2).

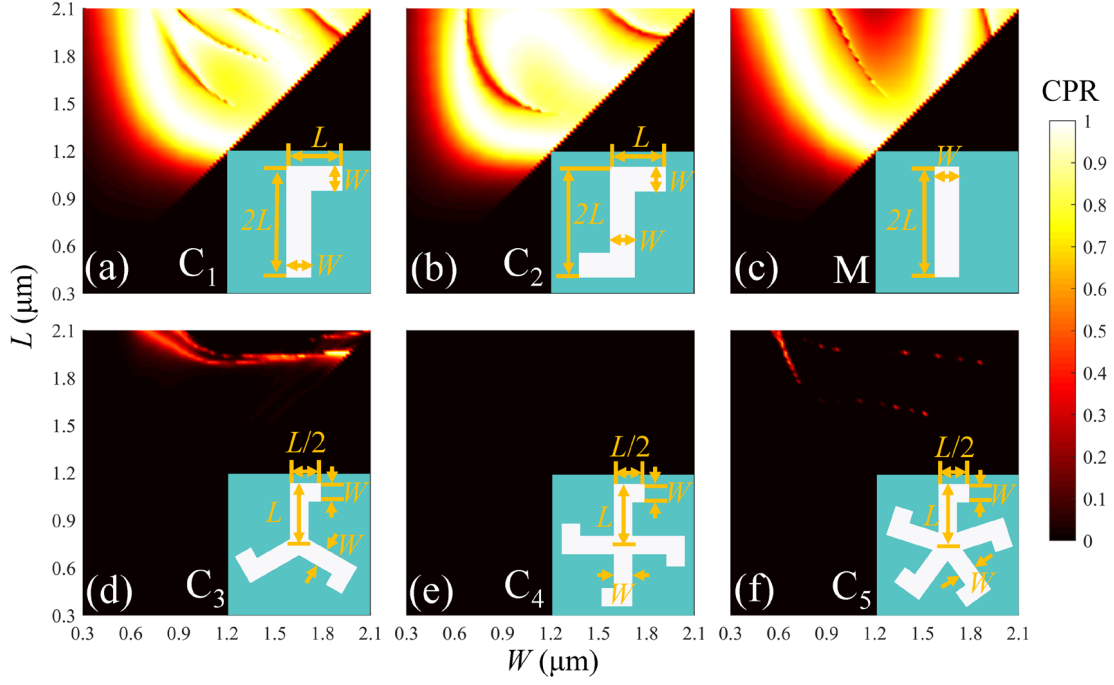

Fig 1. The CPR map as a function of the geometric parameters $L$ and $W$ ($L>W$) for different symmetric meta-atoms at a wavelength of 10.6 μm. Inset: the schematic diagram illustrates meta-atoms with various symmetries, having a height of 6.8 μm and a period of 4.6 μm.

Based on the reasons given above and the fact that an unrestricted meta-atom with C1-symmetry has too many geometric parameters to define, we adopt the constraint of chiral C2-symmetry in the design of the meta-atoms for this paper. The transmission electric field $\vec{E}_t^f$ of the chiral C2-symmetric meta-atoms under forward illumination along the $z$ direction can be expressed as (See Sec. S3 of Supplemental Material for more details [33]):

$$\vec{E}_t^f = \begin{cases} A_0 \left( p_{LL} \vec{s}_L + p_{RL} \vec{s}_R \right) = A_0 \left( A_1 \vec{s}_L + A_2 e^{-j\varphi_2} \vec{s}_R \right), & I_L \\ A_0 \left( p_{LR} \vec{s}_L + p_{RR} \vec{s}_R \right) = A_0 \left( A_2 e^{j\varphi_2} \vec{s}_L + A_1 \vec{s}_R \right), & I_R \end{cases}, \quad (6)$$

$$\begin{cases} A_1 = \dfrac{\varepsilon_0}{2} \sqrt{\left( \chi_{xx} + \chi_{yy} \right)^2} \\ A_2 = \dfrac{\varepsilon_0}{2} \sqrt{\left( \chi_{xx} - \chi_{yy} \right)^2 + \left( 2\chi_{xy} \right)^2} \\ \varphi_1 = 0 \\ \varphi_2 = -\arctan \dfrac{2\chi_{xy}}{\chi_{xx} - \chi_{yy}} \end{cases}. \quad (7)$$

# 3. Design of meta-atoms with H-like shape

Under CP light illumination, the PB phase $\phi_{PB}$ (or so-called geometric phase) for an anisotropic meta-atom rotated around its geometric center by an angle $\theta$ can be expressed as [34]:

$$\phi_{PB} = \arg\left[\cos^2\left(\frac{\delta}{2}\right) + \sin^2\left(\frac{\delta}{2}\right)e^{-2j\sigma\theta}\right], \tag{8}$$

where $\delta$ is the retardation between the long and short axes, which can be calculated as the phase difference between *x*- and *y*-polarized electric field incidence, that is $\delta = \arg(E_{yt}) - \arg(E_{xt})$. $\sigma$ equal to -1/1 for RCP/LCP illumination. When the anisotropic meta-atom meets the condition of a half-wave plate ($\delta=\pi$), the ideal PB phase is:

$$\phi_{PB} = -2\sigma\theta. \tag{9}$$

By combining the PB phase modulation and the phase delay of the planar chiral meta-atom, the Jones matrix in circular basis can be expressed as:

$$J(\theta) = \begin{bmatrix} t_{RR}e^{j\phi_{\chi^{RR}}} & t_{RL}e^{j(\phi_{\chi^{RL}}+\phi_{PB})} \\ t_{LR}e^{j(\phi_{\chi^{LR}}+\phi_{PB})} & t_{LL}e^{j\phi_{\chi^{LL}}} \end{bmatrix}, \tag{10}$$

where $t_{RR}$, $t_{RL}$, $t_{LR}$ and $t_{LL}$ are the transmission amplitude, and $\phi_{\chi^{RR}}$, $\phi_{\chi^{RL}}$, $\phi_{\chi^{LR}}$ and $\phi_{\chi^{LL}}$ are the phase response of the chiral meta-atom, respectively. It should be known that the phase response of the chiral meta-atom can be expressed as the combination of the phase delay from the guided mode propagation $\phi_P$ and the additional phase delay caused by the electric dipole radiation in the chiral meta-atom $\pm\varphi_{1(2)}$ in Eq. (2), that is:

$$\begin{cases} \phi_{\chi^{RR}} = \phi_P - \varphi_1 \\ \phi_{\chi^{RL}} = \phi_P - \varphi_2 \\ \phi_{\chi^{LR}} = \phi_P + \varphi_2 \\ \phi_{\chi^{LL}} = \phi_P + \varphi_1 \end{cases}. \tag{11}$$

From Eq. (11), it can be seen that the chiral meta-atom is spin-decoupled when either $\varphi_1$ or $\varphi_2$ is non-zero. The independent total phase delay can be described by:

$$\begin{cases} \phi_{RR} = \phi_{\chi^{RR}} = \phi_P - \varphi_1 \\ \phi_{RL} = \phi_{\chi^{RL}} + \phi_{PB} = \phi_P - \varphi_2 + \phi_{PB} \\ \phi_{LR} = \phi_{\chi^{LR}} + \phi_{PB} = \phi_P + \varphi_2 + \phi_{PB} \\ \phi_{LL} = \phi_{\chi^{LL}} = \phi_P + \varphi_1 \end{cases} \quad (12)$$

Eq. (12) shows that the PB phase is only applicable to the cross-polarization. Additionally, as indicated in Eq. (7), for chiral C2-symmetric meta-atoms, $\varphi_1$ is zero. This leads to the co-polarization without spin-decoupled functionality. Therefore, the cross-polarization component plays a central role in independently modulating spin light. In order to improve the modulation efficiency and the signal-to-noise ratio, it is significant to minimize the circular dichroism [21, 25] of the chiral meta-atoms and promote the cross-polarization ratio.

As discussed in Sec. 2 and illustrated in Fig. S1(b), adding small rectangles to both sides of a nanofin to break the mirror symmetry can excite induced current flow in the orthogonal direction and generate distinct phase shifts for different CP light. To enhance this induced current flow and increase the number of optimizable parameters, we introduce two small rectangles on either side of a nanofin, forming a H-like shaped meta-atom with 8 variable geometric parameters, as depicted in Fig. 2(a). The basic shape adheres to the chiral C2-symmetry. As is well-known, longer operating wavelengths correspond to larger meta-atom sizes, which are generally easier to fabricate. 10.6 μm corresponds to the output wavelength of $CO_2$ lasers [35] and also serves as the effective wavelength for $SF_6$ gas detection [36]. Therefore, we set the working wavelength to be 10.6 μm in this paper. Please note that the design method presented here can be easily extended to other wavelengths, ensuring its applicability and versatility. In the numerical simulations, the meta-atoms are chosen to be amorphous Si nano-antennas ($n_{Si}$=3.7400) on a $BaF_2$ substrate ($n_{BaF2}$=1.4069) with height $H$=6.8 μm and period $P$=4.6 μm, which can be obtained by a state-of-the–art standard nanofabrication technology [35]. The finite-difference time-domain (FDTD) method (FDTD Solutions, Lumerical, Canada) is used for numerical calculations here (See Sec. S4 of Supplemental Material for more details [33]). To efficiently explore the vast design space and handle numerous variables of the meta-atom, an optimization algorithm [37,38] is utilized to expedite the structure screening process. To avoid converging to local optimal solutions, an adaptive genetic algorithm is implemented in the optimization process (See Sec. S5 of Supplemental Material for more details [33]). Fig. S5 illustrates the schematic diagram of 10 optimized meta-atoms, while Table S1 provides the respective parameters for each of them. Table S2 presents the specific phase, PCE, CPR and transmission values.

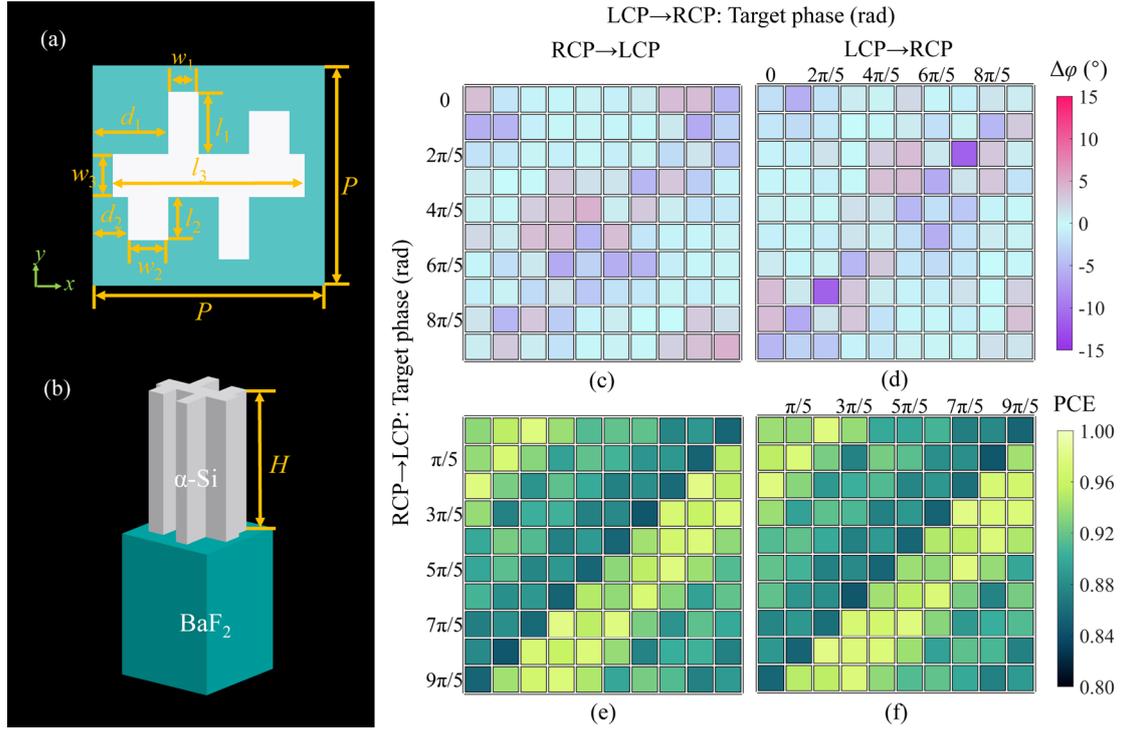

Fig. 2. (a)(b) The schematic top/side view of the H-like shaped meta-atom. Geometric parameters are marked in yellow arrows. The meta-atom height $H$=6.8 μm, period $P$=4.6 μm. (c)(d) The phase difference ($\Delta\varphi$) between reality and target in the phase library (optimized through the combination of meta-atoms and their enantiomers). (e)(f) The map of the polarization conversion efficiency (PCE) for the chosen meta-atoms. The color bar is normalized by Fig. S6.

By combining 10 optimized meta-atoms and their PB phase delay, a spin-decoupled phase library is built spanning the full $2\pi$ range with an interval of $\pi/5$. Figs. S6(a)(b) illustrate the presence of phase difference between the reality and the target in the phase library. We attribute these deviations to two factors: (i) The first factor arises from the PB phase delay offset caused by the mismatch of retardation. According to Eq. (8), when the retardation of the meta-atom does not precisely satisfy the condition of $\delta=\pi$, there will be a PB phase delay offset compared to the ideal $\phi_{PB}$ obtained in Eq. (9) [30]. Fig. S7(a) displays the actual retardations $\delta$ of the optimized meta-atoms, revealing that they are not exactly equal to $\pi$. Consequently, certain offsets exist. Figs. S7(b)(c) depict the offset when $\delta=0.8\pi$ as an example. (ii) The second factor is related with the variations in the propagation phase delay $\phi_p$ of the guide mode, as described in Eq. (12). When the meta-atom is rotated, the relative position between adjacent meta-atoms changes (Fig. S7(d)), influencing the propagation phase delay $\phi_p$.

Interestingly, the mirror symmetry operation on the C2-symmetric chiral meta-atoms yields a corresponding opposite phase response. Specifically, if the relation before the operation is $\varphi_{RL} = \varphi_a$ and $\varphi_{LR} = \varphi_b$, then it becomes $\varphi_{RL} = \varphi_b$ and $\varphi_{LR} = \varphi_a$ after the operation (See Sec. S3 of Supplemental Material for more details

[33]). Based on this symmetry characteristic, the enantiomer of the 10 optimized meta-atoms can also fulfill the requirements of the phase delay library, as shown in Figs. S6(e)(f). Through meticulous comparison of the phase delay between the optimized meta-atoms and their enantiomers, we select those exhibiting minimal deviation from the target phase delay and put them into the phase library. Figs. 2(c)(d) display the phase difference of the optimized phase library. Obviously, the shown deviation is smaller than that either from the optimized meta-atoms (Figs. S6(a)(b)) or their enantiomers (Figs. S6(e)(f)). Notably, the PCEs depicted in Figs. 2(e)(f) are larger than 85% in the whole phase library, indicating high efficiency of the meta-atoms within the library.

## 4. Polarization-insensitive metalens

To demonstrate the high modulation efficiency and signal-to-noise ratio of the chiral C2-symmetry meta-atoms with H-like shape, we utilize the phase library built in Sec. 3 to construct two polarization-insensitive metalenses with numerical aperture (NA) values of 0.80 and 0.40. As commonly known, any polarization can be decomposed into a pair of orthogonal polarization (e.g., *x*- and *y*-polarization or RCP and LCP). When the phase response of RCP and LCP light are identical, it guarantees polarization-insensitive characters. As seen in Table S2, the 1st-5th H-like shaped meta-atoms show the same phase response for RCP and LCP incidence. Thus, they are selected to construct the metalens as the schematic diagram of Fig. 3(a).

For RCP/LCP illumination, the phase at position (*x*, *y*) in space should satisfy the spherical focusing equation [4,5]:

$$\varphi_{LR/RL}(x, y, \lambda) = -\frac{2\pi}{\lambda}\left(\sqrt{x^2 + y^2 + f^2} - f\right), \quad (13)$$

where $\lambda$ is the design wavelength and *f* is the focal length. Both a high NA metalens (The radius *R*=161 μm, focus length *f*=120 μm, NA=0.80) and a low NA metalens (The radius *R*=161 μm, focus length *f*=370 μm, NA=0.40) are built in this section. Figs. 3(b)(c) show the far field intensity profiles of the two metalenses, while Table S3 provides the specific information regarding the actual focus length, full width at half maximum (FWHM) and focusing efficiency. The actual focus length difference between two polarizations is observed to be less than 1.7% (Δ*f/f*), indicating polarization-insensitive characteristics. Furthermore, FWHM approaches the diffraction limit (λ/2NA, 6.63 μm for NA=0.80 and 13.25 μm for NA=0.40), and the focusing efficiency reaches 76%, demonstrating high-quality focusing performance. Table S4 presents a comparison of the focusing efficiency between this paper and those previously reported monochromatic polarization-insensitive metalenses operating in the long-wave infrared band. The results demonstrate highly competitive performance in terms of focusing efficiency. Notably, our metalens achieves diffraction-limited focusing with high focusing efficiency, without the need for a significantly increased aspect ratio of meta-atoms. Moreover, the metalens maintains its competitive focusing efficiency even when increasing the NA from 0.40 to 0.80.

As observed from Figs. 3(b)(c) and Table S3, the intensity profiles at the focus spot in the *x-y* plane exhibit asymmetry for both *x*- and *y*-polarizations. It is more obvious NA=0.80. It can be attributed to the radiation pattern of the electric dipole when excited by linear polarization, which deviates from a perfect spherical wave. The radiation power is highest in the direction perpendicular to the dipole moment, and it decreases as the radiation gets closer to the dipole moment. Generally, the pattern of the dipole radiation in the linear base can be described as an ellipsoidal wave. The long axis of the ellipsoid is oriented perpendicular to the dipole moment, while the short axis is parallel to the dipole moment. The Huygens-Fresnel diffraction principle states that the intensity distribution at the focal point is formed through a coherent superposition of the sub-wavefronts emitted by each meta-atom on the metalens [39]. The linearly polarized incident light excites dipole radiation with the moment parallel to the meta-surfaces and along with the polarization direction. Therefore, at the edge of the metalens, the angle between the dipole radiation and the moment direction becomes smaller at a higher NA. This leads to a more non-uniform intensity distribution pattern at the focal point. As a result, the focused spot of the metalens appears more elliptical for NA=0.80 than 0.40. (See Sec. S8 of Supplemental Material for more details [33]).

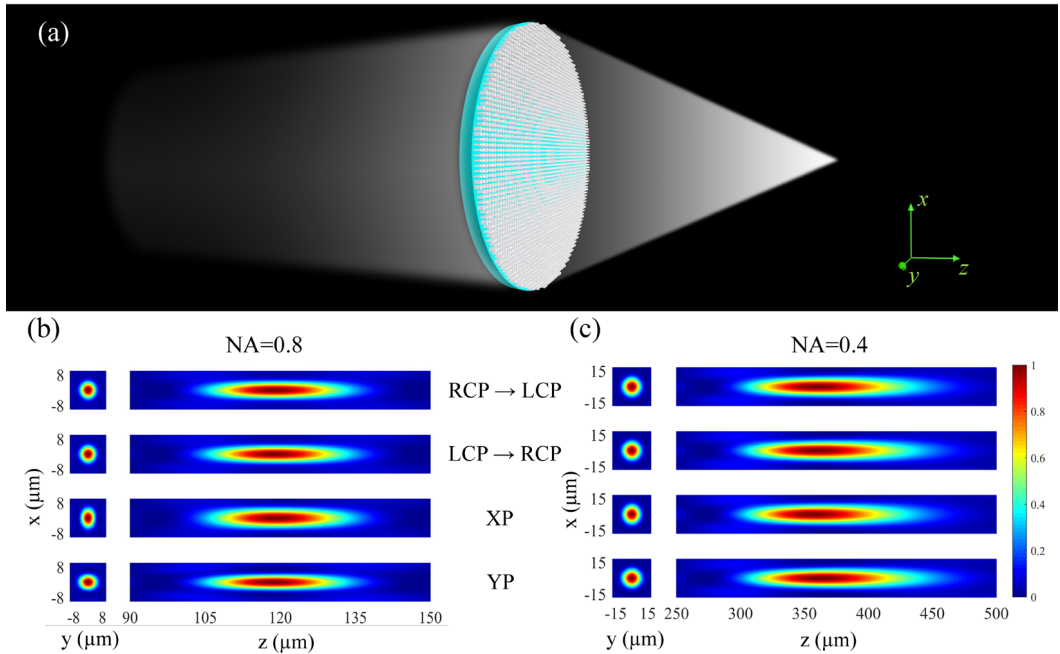

Fig. 3. (a) The schematic diagram illustrates a polarization-insensitive metalens with a radius of 161 μm. (b)/(c) The corresponding intensity profiles in the *x-y* and *x-z* plane with a focal length *f*=120 μm/370 μm for different polarization illuminations. XP: *x*-polarization; YP: *y*-polarization.

## 5. Chiral virtual-moving metalens array

The virtual-moving metalens array has great potential in various applications, including high efficiency and high spatial resolution light field cameras [14,32], vector visual cryptography [20], multiphoton quantum sources [40] and so forth. In this section, a chiral virtual-moving metalens array is designed to demonstrate the high efficiency spin-decoupled multiplexing function of the H-like shaped meta-atoms. The metalens array can efficiently focus different circularly polarized light into distinct spatial positions. The phase profiles responsible for the lateral separation of the focus spots are expressed as [14]:

$$\begin{cases} \varphi_{LR}(x,y,\lambda) = -\frac{2\pi}{\lambda}\left(\sqrt{x^2+y^2+f^2}-f\right) \\ \varphi_{RL}(x,y,\lambda) = -\frac{2\pi}{\lambda}\left(\sqrt{(x+R)^2+(y+R)^2+f^2}-f\right) \end{cases}, \quad (14)$$

where, $R$ is the radius of a single metalens. The above equation indicates that both RCP and LCP will converge at the same focal plane along the $z$-axis, while exhibit a displacement of radius $R$ in the $x$-$y$ plane. Fig. 4(a) depicts the schematic diagram of the 2×2 chiral virtual-moving metalens array, where the side length and focal length are designed as 138 μm and 120 μm, respectively. In Figs. 4(b)-(e), the far field intensity profiles are presented for different polarization illuminations, and the detailed focusing performance of the metalens array is provided in Table S5. Notably, the focusing efficiency for both RCP and LCP reach 84%, with the FWHM of the focus spot closely approaching the diffraction limit (8.41 μm for NA=0.63).

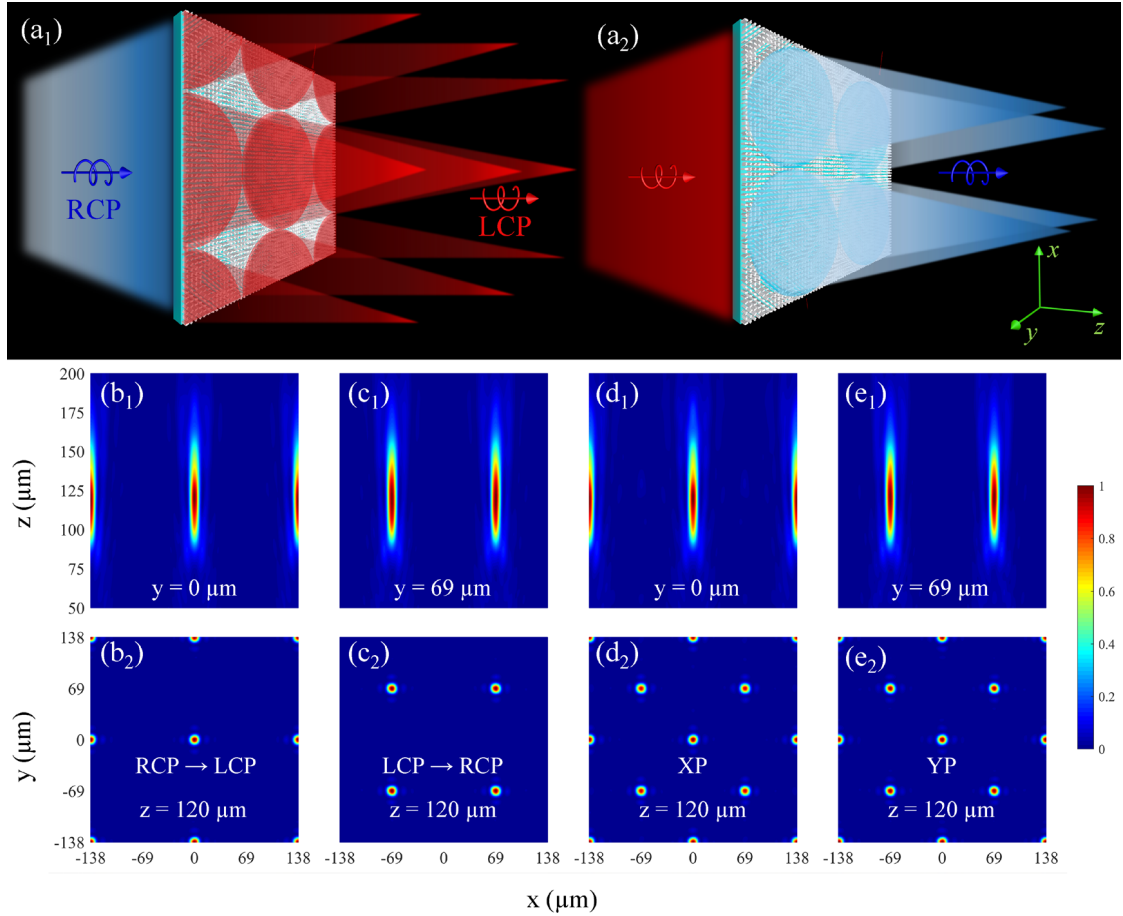

Fig. 4. (a) The schematic diagram of the 2×2 chiral virtual-moving metalens array. The side length and focus length of the metalens is 138 μm and 120 μm. The blue/red arrows represent RCP/LCP light. (b)-(e) Corresponding intensity profiles in the *x-y* and *x-z* plane for different incident polarizations.

## 6. Conclusion

The design in this paper incorporates chiral meta-atoms and their PB phase delay to achieve independent control over spin light. Based on the Jones calculus, our findings demonstrate that chiral C2-symmertic meta-atoms have a greater likelihood of achieving a high cross-polarization ratio. Employing an adaptive genetic algorithm, 10 chiral C2-symmetric meta-atoms with a H-like shape was designed, allowing for the coverage of a full 2π phase span with intervals of π/5 for both spin light components through their rotation. To minimize deviations from the target phase delays, the phase library is updated by integrating the enantiomer of optimized meta-atoms. Utilizing the phase library, we designed a polarization-insensitive metalens and a virtual-moving metalens array to demonstrate their spin-decoupled functionality and high efficiency modulation. Our work has the potential to inspire further advancements in the field of chiral optics and polarization multiplexing metasurfaces, opening up exciting opportunities for future applications.

**Acknowledgements.** This work was funded by National Natural Science Foundation of China (NSFC) (Grant No. 12274148) and Natural Science Foundation of Guangdong province (Grant No. 2021A1515010286). K. S. W. acknowledges the funding support of the Research Grants Council of Hong Kong (Grant No 16304722).

**Disclosures.** The authors declare no conflicts of interest.

**Data availability.** The data that support the findings of this study are available on request from the corresponding author on reasonable request.

# Supplemental Material for: High efficiency spin-decoupled modulation using chiral C2-symmetric meta-atoms

**Author:** Haohan Chen,[1] Jiepeng Wu,[1] Minglei He,[1] Hao Wang,[1] Xinen Wu,[1] Kezhou Fan,[2] Haiying Liu,[1] Qiang Li,[1] Lijun Wu,[1*] and Kam Sing Wong[2]

[1]*Guangdong Provincial Key Laboratory of Nanophotonic Functional Materials and Devices, School of Information and Optoelectronic Science and Engineering, South China Normal University, Guangzhou 510006, China*

[2]*Department of Physics and William Mong Institute of Nano Science and Technology, The Hong Kong University of Science and Technology, Hong Kong 999077, China*

*Corresponding author: ljwu@scnu.edu.cn

## Contents

# S1. Derivation of the radiation electric field

The circularly polarized (CP) electric field $\vec{E}_{\sigma i}$ illuminating along the z-axis (vertical x-y plane) can be expressed as:

$$\vec{E}_{\sigma i} = \begin{bmatrix} E_{xi} \\ E_{yi} \end{bmatrix} e^{j(kz-\omega t)} = \frac{1}{\sqrt{2}} \begin{bmatrix} 1 \\ \sigma j \end{bmatrix} e^{j(kz-\omega t)}, \tag{1}$$

where σ = +1 and -1 represent left-circular polarization (LCP) and right-circular polarization (RCP), respectively. The components $\vec{E}_{xi}$ and $\vec{E}_{yi}$ correspond to the x- and y- components of the incident electric field. $j$ denotes the imaginary symbol of the complex number, while $k$ represents the wave vector in vacuum and $\omega$ signifies the frequency of light.

The electric field response of the meta-atom can be modeled as an electric dipole $\vec{p}$. Normally, a meta-atom can only excite an electric dipole, leading to the polarization vector $\vec{P} = N\vec{p}$, where $N = 1$. The electric dipole moment of an anisotropic meta-atom can be written as:

$$\begin{bmatrix} p_x \\ p_y \end{bmatrix} = \begin{bmatrix} P_x \\ P_y \end{bmatrix} = \varepsilon_0 \begin{bmatrix} \chi_{xx} & \chi_{xy} \\ \chi_{yx} & \chi_{yy} \end{bmatrix} \begin{bmatrix} E_{xi} \\ E_{yi} \end{bmatrix} = \frac{\varepsilon_0}{\sqrt{2}} \begin{bmatrix} \chi_{xx} & \chi_{xy} \\ \chi_{yx} & \chi_{yy} \end{bmatrix} \begin{bmatrix} 1 \\ \sigma j \end{bmatrix}, \tag{2}$$

where $\vec{p} = p_x \vec{e}_x + p_y \vec{e}_y$ is the electric dipole moment and $\vec{P} = P_x \vec{e}_x + P_y \vec{e}_y$ is the polarization vector. $\vec{e}_x$ and $\vec{e}_y$ are the unit vectors along x and y direction, respectively. $\varepsilon_0$ represents the permittivity in vacuum. $\chi_{xx}$, $\chi_{xy}$, $\chi_{yx}$ and $\chi_{yy}$ correspond to the respective components of the electric polarizability tensor.

Changing from the Cartesian base to the circular base, the new base follow [29]

$$\begin{bmatrix} \vec{s}_L \\ \vec{s}_R \end{bmatrix} = \frac{1}{\sqrt{2}} \begin{bmatrix} 1 & j \\ 1 & -j \end{bmatrix} \begin{bmatrix} \vec{e}_x \\ \vec{e}_y \end{bmatrix} = \frac{1}{\sqrt{2}} \begin{bmatrix} \vec{e}_x + j\vec{e}_y \\ \vec{e}_x - j\vec{e}_y \end{bmatrix}, \tag{3}$$

where $\vec{s}_L$ and $\vec{s}_R$ are the unit vectors of the LCP and RCP, respectively. The electric dipole moment can be transformed as $\vec{p} = p_L \vec{s}_L + p_R \vec{s}_R$ [6], that is:

$$\begin{aligned} \begin{bmatrix} p_L \\ p_R \end{bmatrix} &= \varepsilon_0 \begin{bmatrix} \chi_{LL} & \chi_{LR} \\ \chi_{RL} & \chi_{RR} \end{bmatrix} \begin{bmatrix} E_{+i} \\ E_{-i} \end{bmatrix} \\ &= \frac{\varepsilon_0}{2} \begin{bmatrix} \chi_{xx} + \chi_{yy} + j(\chi_{xy} - \chi_{yx}) & \chi_{xx} - \chi_{yy} - j(\chi_{xy} + \chi_{yx}) \\ \chi_{xx} - \chi_{yy} + j(\chi_{xy} + \chi_{yx}) & \chi_{xx} + \chi_{yy} - j(\chi_{xy} - \chi_{yx}) \end{bmatrix} \begin{bmatrix} E_{+i} \\ E_{-i} \end{bmatrix}. \end{aligned} \tag{4}$$

The $\chi_{LL}$, $\chi_{LR}$, $\chi_{RL}$ and $\chi_{RR}$ are the electric polarizability tensor in circular base. The formal equation can be further simplified as:

$$\begin{bmatrix} p_L \\ p_R \end{bmatrix} = \begin{bmatrix} A_1 e^{j\varphi_1} & A_2 e^{j\varphi_2} \\ A_2 e^{-j\varphi_2} & A_1 e^{-j\varphi_1} \end{bmatrix} \begin{bmatrix} E_{+i} \\ E_{-i} \end{bmatrix} = \begin{bmatrix} p_{LL} & p_{LR} \\ p_{RL} & p_{RR} \end{bmatrix} \begin{bmatrix} s_L \\ s_R \end{bmatrix}, \quad (5)$$

$$\begin{cases} A_1 = \dfrac{\varepsilon_0}{2} \sqrt{(\chi_{xx} + \chi_{yy})^2 + (\chi_{xy} - \chi_{yx})^2} \\ A_2 = \dfrac{\varepsilon_0}{2} \sqrt{(\chi_{xx} - \chi_{yy})^2 + (\chi_{xy} + \chi_{yx})^2} \\ \varphi_1 = \arctan \dfrac{\chi_{xy} - \chi_{yx}}{\chi_{xx} + \chi_{yy}} \\ \varphi_2 = -\arctan \dfrac{\chi_{xy} + \chi_{yx}}{\chi_{xx} - \chi_{yy}} \end{cases} \quad (6)$$

In the notations $p_{LL}$, $p_{LR}$, $p_{RL}$ and $p_{RR}$, the left-hand side subscripts indicate the chirality of the electric dipole radiation field, while the right-hand side subscripts represent the chirality of the incident electric field. For instance, $p_{LR}$ represents the electric dipole moment of the LCP component of the electric dipole radiation field when it is illuminated by RCP light. All the same subscripts in the whole paper represent the same meaning unless otherwise specified.

Therefore, the expression for the transmission electric field can be given as follows:

$$\vec{E}_t = \begin{cases} A_0 \left[ p_{LL} \vec{s}_L + p_{RL} \vec{s}_R - \dfrac{1}{\sqrt{2}} \sin\xi (p_{LL} + p_{RL}) \vec{e}_{R_0} \right], & I_L \\ A_0 \left[ p_{LR} \vec{s}_L + p_{RR} \vec{s}_R - \dfrac{1}{\sqrt{2}} \sin\xi (p_{LR} + p_{RR}) \vec{e}_{R_0} \right], & I_R \end{cases} \quad (7)$$

where $A_0 = \dfrac{-\omega^2}{4\pi\varepsilon_0 c^2 R_0} e^{jkR_0}$, $R_0$ is the distance between the observation point and the electric dipole and $c$ is the velocity of light in vacuum. $I_L$ and $I_R$ represent RCP and LCP incidences, respectively. The spatial unit vector $\vec{e}_{R_0}$ is defined as $\vec{e}_{R_0} = \sin\xi \cos\theta \vec{e}_x + \sin\xi \sin\theta \vec{e}_y + \cos\xi \vec{e}_z$, where $\theta$ represents the angle between the projection of the line from the observation point to the origin of the coordinates in the x-y plane and the x-axis, and $\xi$ is the angle between the line from the observation point to the origin of the coordinates and $z=0$. A schematic diagram is available in Fig. S1(c).

When CP light is illuminating, the radiation electric field, which is perpendicular to the x-y plane (i.e., $\xi=0$), can be simplified as

$$\vec{E}_t = \begin{cases} A_0\left(p_{LL}\vec{s}_L + p_{RL}\vec{s}_R\right) = A_0\left(A_1 e^{j\varphi_1}\vec{s}_L + A_2 e^{-j\varphi_2}\vec{s}_R\right), & I_L \\ A_0\left(p_{LR}\vec{s}_L + p_{RR}\vec{s}_R\right) = A_0\left(A_2 e^{j\varphi_2}\vec{s}_L + A_1 e^{-j\varphi_1}\vec{s}_R\right), & I_R \end{cases}. \quad (8)$$

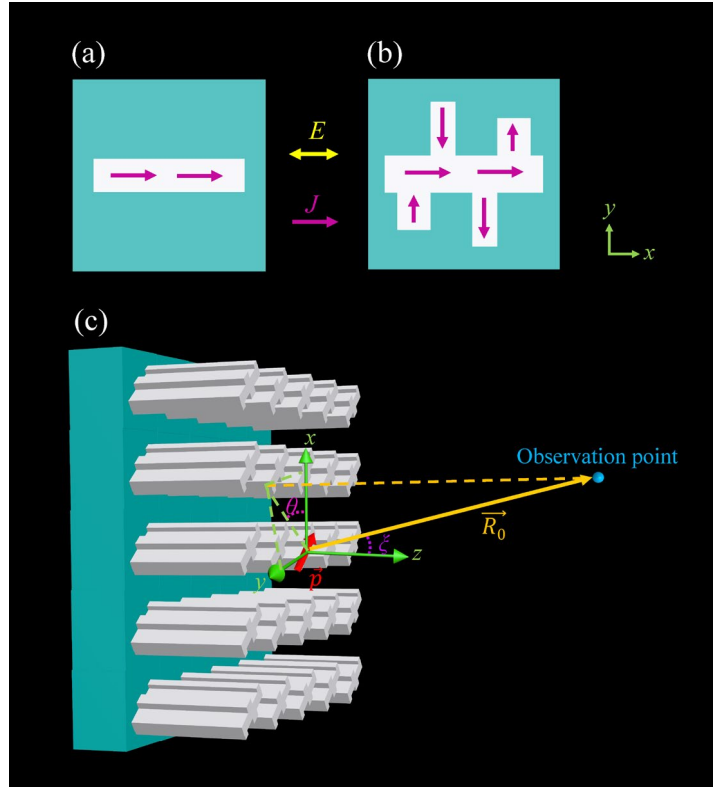

FIG. S1. (a)/(b) The top view of the nanofin/ the H-like shaped meta-atom, with the purple arrows illustrating the current flow induced by the *x*-polarized electric field (the yellow arrow). (c) A schematic diagram depicting the identification of coordinates.

## S2. Derivation of CPR with different symmetry

The cross-polarization ratio (CPR), as defined in Eqs. (3) and (4) in the main text, can be expressed as follow:

$$CPR = \frac{A_2^2}{A_1^2 + A_2^2} = \frac{\left(\chi_{xx} - \chi_{yy}\right)^2 + \left(\chi_{xy} + \chi_{yx}\right)^2}{\left(\chi_{xx} + \chi_{yy}\right)^2 + \left(\chi_{xy} - \chi_{yx}\right)^2 + \left(\chi_{xx} - \chi_{yy}\right)^2 + \left(\chi_{xy} + \chi_{yx}\right)^2}. \quad (9)$$

Considering the forward illumination along the *z* direction, the Jones matric of the electric polarizability tensor $\hat{T}_{lin}^f$ can be expressed as [29]:

$$\hat{T}_{lin}^f = \begin{bmatrix} \chi_{xx} & \chi_{xy} \\ \chi_{yx} & \chi_{yy} \end{bmatrix}, \quad (10)$$

where the superscript *f* represents forward propagation and the subscript *lin* represents the linear orthogonal base. For backward illumination along the *-z* direction, the electric

polarizability tensor $\hat{T}_{lin}^{b}$ can be described as:

$$\hat{T}_{lin}^{b} = \begin{bmatrix} \chi_{xx} & -\chi_{yx} \\ -\chi_{xy} & \chi_{yy} \end{bmatrix}, \quad (11)$$

where superscript $b$ represents backward propagation.

The transformation matrix $D_{\beta}$, which describes the rotation of the meta-atom by an angle $\beta$ along the $z$-axis, is given by:

$$D_{\beta} = \begin{bmatrix} \cos(\beta) & \sin(\beta) \\ -\sin(\beta) & \cos(\beta) \end{bmatrix}. \quad (12)$$

The transformation matrices for mirror symmetry along the $x$-$z$ plane $M_x$ and $y$-$z$ plane $M_y$ can be expressed as:

$$M_x = \begin{bmatrix} 1 & 0 \\ 0 & -1 \end{bmatrix}, M_y = \begin{bmatrix} -1 & 0 \\ 0 & 1 \end{bmatrix}. \quad (13)$$

The electric polarizability tensor for the meta-atom with mirror symmetry along the $x$-$z$ plane $\hat{T}_{M_x}^{f}$ or $y$-$z$ plane $\hat{T}_{M_y}^{f}$ follows the relation as:

$$\begin{cases} M_x^{-1} \hat{T}_{M_x}^{f} M_x = \hat{T}_{M_x}^{f} \\ M_y^{-1} \hat{T}_{M_y}^{f} M_y = \hat{T}_{M_y}^{f} \end{cases}. \quad (14)$$

From Eq. (14), it can be concluded that $\chi_{xy} = \chi_{yx} = 0$ and $\chi_{xx} \neq \chi_{yy} \neq 0$. Therefore, for the meta-atom with mirror symmetry, the electric polarizability tensor $\hat{T}_{M}^{f}$ can be expressed as:

$$\hat{T}_{M}^{f} = \hat{T}_{M_x}^{f} = \hat{T}_{M_y}^{f} = \begin{bmatrix} \chi_{xx} & 0 \\ 0 & \chi_{yy} \end{bmatrix}. \quad (15)$$

The electric polarizability tensor of the meta-atom with mirror symmetry at a certain angle along the $x$-$z$ plane can be transformed to the form described in Eq. (15) using Eq. (12) and (13). By combining Eq. (9) and (15), the CPR of the mirror symmetric meta-atom can be described as:

$$CPR_M = \frac{A_2^2}{A_1^2 + A_2^2} = \frac{(\chi_{xx} - \chi_{yy})^2}{(\chi_{xx} + \chi_{yy})^2 + (\chi_{xx} - \chi_{yy})^2}. \quad (16)$$

Since $\chi_{xx} \neq \chi_{yy} \neq \chi_{xy} \neq \chi_{yx} \neq 0$ for the C1-symmetric meta-atom without mirror symmetry, the electric polarizability tensor $\hat{T}_{C_1}^{f}$ can be expressed as:

$$\hat{T}_{C_1}^f = \hat{T}_{lin}^f = \begin{bmatrix} \chi_{xx} & \chi_{xy} \\ \chi_{yx} & \chi_{yy} \end{bmatrix}. \tag{17}$$

Therefore, the CPR of the C1-symmetric chiral meta-atom can be described as:

$$CPR_{C_1} = \frac{A_2^2}{A_1^2 + A_2^2} = \frac{\left(\chi_{xx} - \chi_{yy}\right)^2 + \left(\chi_{xy} + \chi_{yx}\right)^2}{\left(\chi_{xx} + \chi_{yy}\right)^2 + \left(\chi_{xy} - \chi_{yx}\right)^2 + \left(\chi_{xx} - \chi_{yy}\right)^2 + \left(\chi_{xy} + \chi_{yx}\right)^2}. \tag{18}$$

The C2-symmetric meta-atom without mirror symmetry not only coincides with itself by rotating $\pi$ angle along the z-axis, but also coincides with the backward structure after mirror symmetry along the x-z or y-z plane. The electric polarizability tensor of the C2-symmetric chiral meta-atom satisfies the following relation:

$$\begin{cases} D_\pi^{-1} \hat{T}_{C_2}^f D_\pi = \hat{T}_{C_2}^f \\ M_x^{-1} \hat{T}_{C_2}^f M_x = \hat{T}_{C_2}^b \end{cases}. \tag{19}$$

By combining Eqs. (10)-(12) and (19), it can be observed that $\chi_{xy} = \chi_{yx} \neq 0$ and $\chi_{xx} \neq \chi_{yy} \neq 0$. Therefore, for the C2-symmetric meta-atom without mirror symmetry, the electric polarizability tensor $\hat{T}_{C_2}^f$ can be expressed as:

$$\hat{T}_{C_2}^f = \begin{bmatrix} \chi_{xx} & \chi_{xy} \\ \chi_{xy} & \chi_{yy} \end{bmatrix}. \tag{20}$$

By combining Eq. (9) and (20), the CPR of the C2-symmetric chiral meta-atom can be described as:

$$CPR_{C_2} = \frac{A_2^2}{A_1^2 + A_2^2} = \frac{\left(\chi_{xx} - \chi_{yy}\right)^2 + \left(2\chi_{xy}\right)^2}{\left(\chi_{xx} + \chi_{yy}\right)^2 + \left(\chi_{xx} - \chi_{yy}\right)^2 + \left(2\chi_{xy}\right)^2}. \tag{21}$$

It is evident from Eqs. (18) and (21) that for C1-symmetric and C2-symmetric chiral meta-atoms, if $\chi_{xy} > \chi_{yx}$, then $CPR_{C_2} > CPR_{C_1}$, and vice versa.

The Cn-symmetric (n>2) meta-atom without mirror symmetry coincides with itself by rotating $2\pi/n$ angle along the z-axis. Therefore, the electric polarizability tensor of the Cn-symmetric chiral meta-atom satisfies the following relation:

$$D_{\frac{2\pi}{n}}^{-1} \hat{T}_{C_n}^f D_{\frac{2\pi}{n}} = \hat{T}_{C_n}^f. \tag{22}$$

By combining Eqs. (10), (12) and (22), it can be derived that $\chi_{xy} = -\chi_{yx}$ and $\chi_{xx} = \chi_{yy}$ when n>2. Therefore, for the Cn-symmetric chiral meta-atom, the electric polarizability tensor $\hat{T}_{C_n}^f$ can be expressed as:

$$\hat{T}_{C_n}^f = \begin{bmatrix} \chi_{xx} & \chi_{xy} \\ -\chi_{xy} & \chi_{xx} \end{bmatrix}. \tag{23}$$

From Eqs. (9) and (23), the CPR of the C$n$-symmetric chiral meta-atom can be described as:

$$CPR_{C_n} = \frac{A_2^2}{A_1^2 + A_2^2} = \frac{0}{(2\chi_{xx})^2 + 0} = 0. \tag{24}$$

Actually, the conclusion of this paper, which is derived based on the dipole approximation, states that the CPR of C$n$-symmetric ($n>2$) meta-atoms is precisely 0 for the dipole radiation component. However, it is important to note that the CPR may not be exactly 0 for higher order multipole radiation components. Nevertheless, since the dipole is the dominant multipole component in antenna radiation, the CPR of the dipole component effectively reflects the CPR of the entire meta-atom to a considerable extent.

According to Eqs. (16), (18), (21) and (24), when $\chi_{xx} \neq \chi_{yy} \neq \chi_{xy} \neq \chi_{yx} \neq 0$, the order of the CPR for meta-atoms with different symmetries is as follows:

$$CPR_{C_1},\ CPR_{C_2} > CPR_M > CPR_{C_n},\ (n > 2). \tag{25}$$

It should be noted that Eq. (25) is valid under the assumption of the same polarization tensor for each symmetric meta-atom, which can be challenging to satisfy in practice. In other words, Eq. (25) implies that the parameter space for achieving high CPR in chiral C1- and C2-symmetric meta-atoms is larger compared to their corresponding mirror symmetric and other higher symmetric meta-atoms. To support this concept, Figs. 1, 2S and 3S illustrate the CPR, PCE and transmittance of different symmetric meta-atoms, respectively. The parameters are simplified to $L$ and $W$ ($L>W$) for a more intuitive comparison. The size of the high CPR region for different symmetric meta-atoms confirms the conclusion of Eq. (25). In order to obtain the highest CPR and reduce the optimization parameters, the design of H-like shaped meta-atoms in this paper should satisfy chiral C2-symmetry.

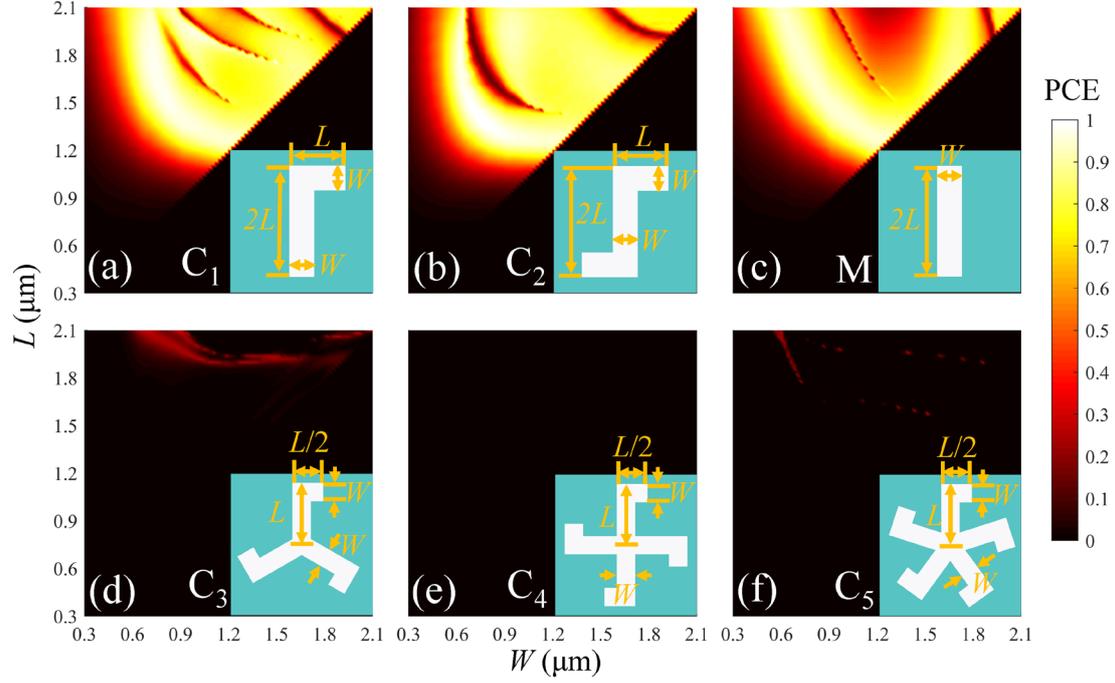

FIG. S2. The PCE map as a function of the geometric parameters $L$ and $W$ ($L>W$) for different symmetric meta-atoms at a wavelength of 10.6 μm. Inset: the schematic diagram illustrates meta-atoms with various symmetries, having a height $H$=6.8 μm and a period $P$=4.6 μm.

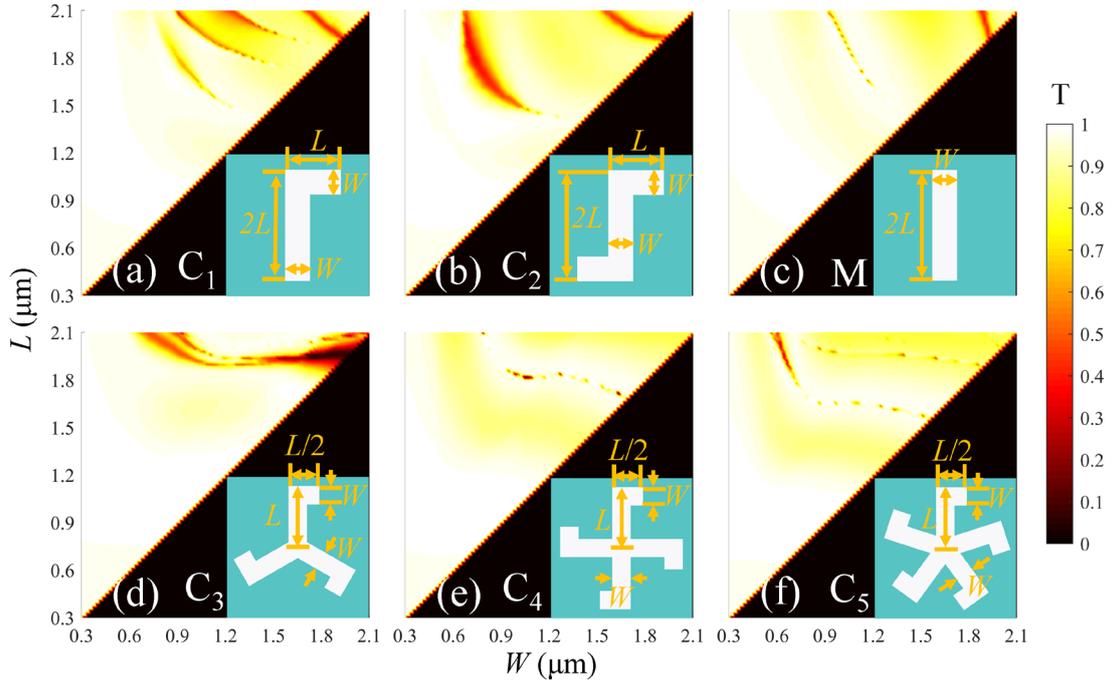

FIG. S3. The Transmittance map as a function of the geometric parameters $L$ and $W$ ($L>W$) for different symmetric meta-atoms at a wavelength of 10.6 μm. Inset: the schematic diagram illustrates meta-atoms with various symmetries, having $H$=6.8 μm and $P$=4.6 μm.

# S3. Derivation of electric field response of chiral C2-symmetric meta-atoms

According to Eq. (20), for chiral C2-symmetric meta-atoms, Eq. (6) can be simplified as:

$$\begin{cases} A_1 = \frac{\varepsilon_0}{2}\sqrt{(\chi_{xx}+\chi_{yy})^2} \\ A_2 = \frac{\varepsilon_0}{2}\sqrt{(\chi_{xx}-\chi_{yy})^2+(2\chi_{xy})^2} \\ \varphi_1 = 0 \\ \varphi_2 = -\arctan\frac{2\chi_{xy}}{\chi_{xx}-\chi_{yy}} \end{cases}. \quad (26)$$

Therefore, from Eqs. (8) and (26), the transmission electric field $\vec{E}_t^f$ of the chiral C2-symmetric meta-atoms under forward illumination along the $z$ direction can be expressed as:

$$\vec{E}_t^f = \begin{cases} A_0(p_{LL}\vec{s}_L + p_{RL}\vec{s}_R) = A_0(A_1\vec{s}_L + A_2 e^{-j\varphi_2}\vec{s}_R), & I_L \\ A_0(p_{LR}\vec{s}_L + p_{RR}\vec{s}_R) = A_0(A_2 e^{j\varphi_2}\vec{s}_L + A_1\vec{s}_R), & I_R \end{cases}. \quad (27)$$

By applying the $x$-mirror symmetry operation $\hat{M}_x$ to the C2-symmetric chiral meta-atoms, and considering Eqs. (13) and (20), the electric polarizability tensor $\hat{T}_{C_2\hat{M}_x}^f$ can be expressed as:

$$\hat{T}_{C_2\hat{M}_x}^f = M_x^{-1}\hat{T}_{C_2}^f M_x = \begin{bmatrix} \chi_{xx} & -\chi_{xy} \\ -\chi_{xy} & \chi_{yy} \end{bmatrix}. \quad (28)$$

Substituting Eq. (28) into Eq. (26) gives

$$\begin{cases} A_{1\hat{M}_x} = \frac{\varepsilon_0}{2}\sqrt{(\chi_{xx}+\chi_{yy})^2} = A_1 \\ A_{2\hat{M}_x} = \frac{\varepsilon_0}{2}\sqrt{(\chi_{xx}-\chi_{yy})^2+(2\chi_{xy})^2} = A_2 \\ \varphi_{1\hat{M}_x} = 0 = -\varphi_1 \\ \varphi_{2\hat{M}_x} = \arctan\frac{2\chi_{xy}}{\chi_{xx}-\chi_{yy}} = -\varphi_2 \end{cases}, \quad (29)$$

Therefore, the expression for the transmission electric field $\vec{E}_{t\hat{M}_x}^f$ of the chiral C2-symmetric meta-atoms after the $x$-mirror symmetry operation $\hat{M}_x$, under forward

illumination along the z direction, can be given as:

$$\vec{E}^f_{t\hat{M}_x} = \begin{cases} A_0\left(A_1\vec{s}_L + A_2 e^{j\varphi_2}\vec{s}_R\right), & I_L \\ A_0\left(A_2 e^{-j\varphi_2}\vec{s}_L + A_1\vec{s}_R\right), & I_R \end{cases}. \tag{30}$$

Comparing Eqs. (27) and (30), it can be observed that the phase response of the C2-symmetry structure after the x-mirror symmetry operation $\hat{M}_x$ is opposite to that before the mirror symmetry operation. In other words, if the phase delay before the mirror symmetry operation is $\varphi_{RL} = \varphi_a$ and $\varphi_{LR} = \varphi_b$, then the phase delay after the mirror symmetry operation is $\varphi_{RL} = \varphi_b$ and $\varphi_{LR} = \varphi_a$. Utilizing this property, we can compare the structures before and after the mirror symmetry operation to identify the structure that is closer to the target phase.

## S4. Numerical simulations

The numerical simulations in this paper utilize the finite-difference time-domain (FDTD) method implemented in the FDTD Solutions software developed by Lumerical, Canada. A mesh size of 130 nm is chosen to ensure the convergence of the results. The meta-atoms investigated in this study are composed of amorphous Si nano-antennas ($n_{Si}$=3.7400) placed on a BaF$_2$ substrate ($n_{BaF2}$=1.4069). These meta-atoms operate at a wavelength of 10.6 μm, with a height of $H$=6.8 μm and a period of $P$=4.6 μm.

The transmission phase, polarization conversion efficiency (PCE) and transmissivity of the unit cell were simulated by the FDTD method. Periodic boundary conditions were set along the x- and y-axis, and perfect matching layers were applied along the z-axis to minimize reflection. A plane wave with circular polarization served as the incident light source. The PCE, which is defined as the ratio of the transmission with opposite chirality to the incidence, can also be expressed as the product of the CPR and transmissivity.

The focusing efficiency, focus length, full width half maximum (FWHM) of the metalens and far field intensity profiles of the metalens were simulated by the FDTD method. Perfect matching layers were implemented along the x, y and z directions. The light source used was a total-field scattered-field (TFSF) with circular or linear polarization. The focusing efficiency is defined as the ratio of the power at the focal spot to the incident power.

## S5. Adaptive genetic algorithm

The genetic algorithm [37, 38] is a well-known global optimization technique renowned for its fast convergence and high accuracy. However, when applied to the optimization of H-like shaped chiral C2-symmetric meta-atoms, which involve 8 parameters and offer significant design flexibility, the simple genetic algorithm often becomes susceptible to local optima. To address this issue, we have adopted an adaptive genetic algorithm that incorporates self-adaptive mutation and crossover operations to optimize the H-like shape meta-atoms. To assess whether the optimization process becomes trapped in local optima, we have introduced a population entropy-based criterion. When the optimization process is identified to be trapped in a local optimum, the adaptive genetic algorithm executes a jump-out operation, enabling escape from the suboptimal solution and facilitating the attainment of global optimization. The flow diagram of the self-adaptive genetic algorithm is presented in Fig. S4. Further details of these operations are described as follows:

**Initialize population**: The initial population comprised 30 randomly generated individuals, with their parameters encoded using the Gray coding technique.

**Update population**: The individuals in the population were evaluated to determine if they satisfied the constraint conditions. If any individual did not meet the conditions, it was regenerated until all individuals satisfied the constraints. The constraint conditions, aimed at ensuring the manufacturability of the optimized meta-atoms and preventing overlap when rotating, included the conditions $(w_1+w_2+d_1+d_2)<(P-0.8\ \mu m)$, $(2\max(l_1, l_2)+w_3)<(P-0.8\ \mu m)$, $(\min(d_1, d_2)+ l_3/2)>(P/2)$, $((P/2-d_1)^2+(w_3/2+ l_1)^2)<(P/2-0.4\ \mu m)^2$ and $((P/2-d_2)^2+(w_3/2+ l_2)^2)<(P/2-0.4\ \mu m)^2$.

**Calculate figure of merit (FOM)**: As mentioned in Sec. 3 of the main text and Supplemental Material Sec. S7, there are additional phase delays caused by the offset of the PB phase and the change in the guide mode propagation phase. To minimize these additional phase delays, the actual phase response of the meta-atoms when rotated at 45 degrees is taken into account in the FOM. The FOM for a single wavelength is defined as:

$$\begin{cases} \text{FOM}_0 = \left| e^{|\Phi-\Phi_0|} + \dfrac{|T - T_0|}{T_0} \right| -1 \\ \text{FOM}_{45} = \left| e^{\left|\Phi_{45}-\Phi_0+\frac{\pi}{2}\sigma\right|} + \dfrac{|T_{45} - T_0|}{T_0} \right| -1 \\ \text{FOM} = \dfrac{1}{0.3(\text{FOM}_0)^2 + 0.3(\text{FOM}_{45})^{1.5} - 0.6} \end{cases}, \qquad (31)$$

where $\Phi$ and $\Phi_0$ are the actual and target phase, respectively, when the meta-atoms are rotated at 0 degrees. $\Phi_{45}$ is the actual phase when the meta-atoms are rotated at 45 degrees, and $\sigma$ takes a value of -1/+1 for RCP/LCP illumination. $T$ and $T_0 =1$ are the actual and target PCE when the meta-atoms are rotated at 0 degrees, respectively. $T_{45}$ =1 is the target PCE when the meta-atoms are rotated at 45 degrees. Therefore, the selection probability of each individual can be determined by

$$P_i = \dfrac{\text{FOM}_i}{\sum\limits_{i=1}^{N} \text{FOM}_i}, \qquad (32)$$

where $P_i$ denotes the selection probability and $FOM_i$ represents the FOM for the $i^{th}$ individual. $N = 30$ corresponds to the total number of individuals in the population.

**Calculate population entropy**: Similar to information entropy, population entropy can be used to quantify the degree of population diversity. The population entropy $H_{pe}$ is defined as

$$H_{pe} = -\sum_{i=1}^{N} P_i \ln(P_i). \tag{33}$$

A higher population entropy indicates a lower degree of population diversity. The maximum population entropy is given by $H_{pe}^{max} = \ln(N)$. If the $H_{pe} > 0.95 H_{pe}^{max}$ for two consecutive iterations, it can be inferred that the optimization process is trapped in a local optimum, necessitating the implementation of a jump-out operation to escape the suboptimal solution.

**Selection operation**: After calculating the selection probability for each individual, the Roulette Wheel method is employed to select the individual with high FOM for next iteration, with a selection rate of 0.5. However, when certain individuals exhibit significantly higher FOM compared to others, these super individuals have a greater probability of being selected, increasing the risk of getting trapped in a local optimum due to reduced population diversity. To mitigate this, a FOM scale transformation is introduced for each individual prior to the selection process, which reduces the selection probability of super individuals and increases the selection probability for individuals with lower FOM. Specifically, the transformed FOM, denoted as FOM', for the $i^{th}$ individual is calculated as follows:

$$\text{FOM}' = \begin{cases} \text{ave(FOM)} + \mu\sqrt{\dfrac{\text{FOM}_i - \text{ave(FOM)}}{\mu}}, & \text{FOM}_i \geq \text{ave(FOM)} \\ \text{ave(FOM)} - \mu\sqrt{\dfrac{\text{ave(FOM)} - \text{FOM}_i}{\mu}}, & \text{FOM}_i < \text{ave(FOM)} \end{cases}, \tag{34}$$

where ave(FOM) represents the average FOM in the population, and $\mu$ is defined as:

$$\begin{cases} \mu = 0.5 \min(\mu_1, \mu_2) \\ \mu_1 = \max(\text{FOM}) - \text{ave(FOM)} \\ \mu_2 = \text{ave(FOM)} - \min(\text{FOM}) \end{cases}, \tag{35}$$

where max(FOM) and min(FOM) are the maximum and minimum FOM in the population, respectively.

**Self-adaptive crossover operation**: In genetic algorithms, the crossover probability $p_c$ plays a crucial role in generating new individuals. A higher $p_c$ accelerates the generation of new individuals but also poses a greater risk of destroying high FOM individuals. Conversely, a smaller $p_c$ leads to a slower search speed. Hence, it is important to strike a balance between the evolutionary direction and search speed by appropriately adjusting the crossover probability. To address this, we employ a self-adaptive crossover operation that assigns a low crossover probability to individuals with high FOM and a high crossover probability to those with low FOM. The self-adaptive crossover probability $p_c$ is defined as:

$$p_c = \begin{cases} p_{c1} - \dfrac{(p_{c1}-p_{c2})(\text{FOM}_j - \text{ave}(\text{FOM}))}{\max(\text{FOM}) - \text{ave}(\text{FOM})}, & \text{FOM}_j \geq \text{ave}(\text{FOM}) \\ p_{c1}, & \text{FOM}_j < \text{ave}(\text{FOM}) \end{cases}, \quad (36)$$

where FOM$_j$ represents the higher FOM value between the two parents selected for the crossover operation. The $p_{c1}$=0.9 and $p_{c2}$=0.3. We applied the multi-point crossover operation, with the ratio of crossover points to the overall number of points in the individuals $p_{cn}$ is set $p_{cn} = p_c^3$.

**Self-adaptive mutation operation**: Increasing the mutation probability $p_m$, akin to the crossover probability $p_c$, expedites the generation of new individuals while simultaneously amplifying the attrition of high FOM individuals. Conversely, reducing the value of $p_m$ diminishes the search speed. Consequently, to mitigate these concerns, we implement a self-adaptive mutation operation, wherein individuals with high FOM are allocated a low mutation probability and those with low FOM are assigned a high mutation probability. The expression denoting the self-adaptive mutation probability, $p_m$, is given as follows:

$$p_m = \begin{cases} p_{m1} - \dfrac{(p_{m1}-p_{m2})(\max(\text{FOM}) - \text{FOM}_k)}{\max(\text{FOM}) - \text{ave}(\text{FOM})}, & \text{FOM}_k \geq \text{ave}(\text{FOM}) \\ p_{m1}, & \text{FOM}_k < \text{ave}(\text{FOM}) \end{cases}, \quad (37)$$

where, FOM$_k$ is the FOM of the individual prepared for the mutation operation. The $p_{m1}$=0.2 and $p_{m2}$=0.01. The ratio of mutation points to the total number of points in the individuals $p_{mn}$ is set $p_{mn} = p_m^2$.

**Jump-out local optima operation**: As previously mentioned, we consider the optimization process to be trapped in local optima if the $H_{pe} > 0.95 H_{pe}^{\max}$ for two consecutive iterations. To address this situation, we employ the jump-out local optima operation, which involves modifying the selection, crossover and mutation operations. For the selection operation, the selection rate is reduced to 0.1. In the crossover operation, the crossover probability $p_c$ is increased to 0.9, and the crossover point ratio $p_{cn}$ is set to 0.5. Regarding the mutation operation, in addition to changing the mutation probability $p_{cm}$ to 0.9, we adopt the bitwise mutation operation. In a population with binary coding, the monotonic coefficient of the $i^{th}$ coded bit can be defined as

$$\mu_i = \left| \dfrac{1}{N} \sum_{j=1}^{N} x_{ji} - 0.5 \right| \times 2, (j=1,2,\cdots,N; i=1,2,\cdots,M), \quad (38)$$

where, the $\mu_i$ is the monotonic coefficient of the $i^{th}$ coded bit. $x_{ij}$ signifies the $i^{th}$ coded bit value of the $j^{th}$ individual in the population. $N$ denotes the total number of individuals in the population and $M$ represents the total number of coding bits of the individual. When all the numbers in the $i^{th}$ bit of the population are exclusively 0 or 1, $\mu_i$=1.

Conversely, if the $i^{th}$ bit of the population consists of an equal number of 0s and 1s, $\mu_i$=0. During the bitwise mutation operation, the mutation probability for the $i^{th}$ coding position of the $j^{th}$ individual is determined as $p_{mji} = \mu_i / 2$.

**Optimization termination**: If the maximum FOM remains unchanged after 5 continuous jump-out local optima operations, it can be considered as the global optimum solution. Otherwise, the algorithm will terminate at the $200^{th}$ generations.

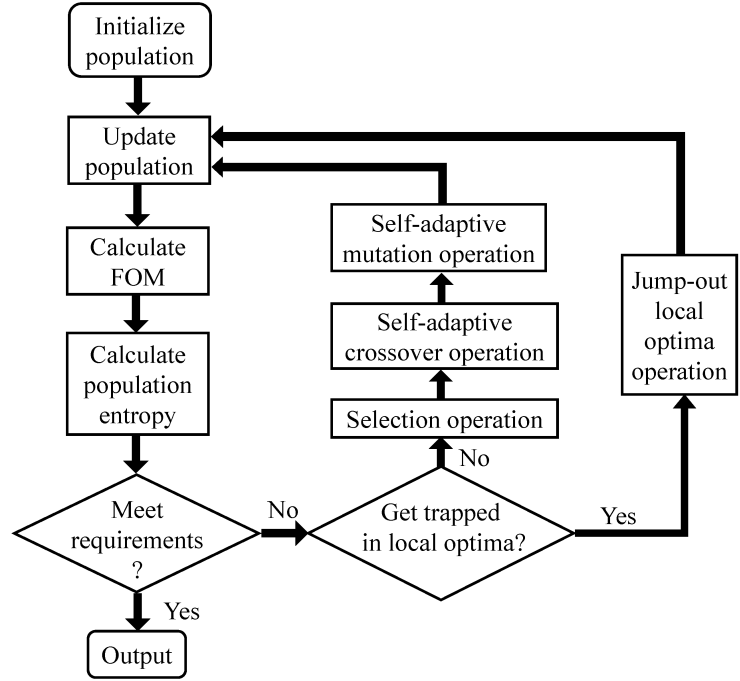

FIG. S4. The flow diagram of the self-adaptive genetic algorithm.

## S6. The optimized meta-atoms and the phase library

TABLE S1. The parameters of 10 optimized meta-atoms with H-like shape.

| No. | W1 (μm) | W2 (μm) | W3 (μm) | L1 (μm) | L2 (μm) | L3 (μm) | d1 (μm) | d2 (μm) |
|---|---|---|---|---|---|---|---|---|
| 1 | 0.758 | 0.461 | 0.680 | 0.427 | 0.291 | 3.444 | 1.768 | 0.628 |
| 2 | 0.411 | 0.469 | 0.886 | 0.713 | 0.460 | 2.839 | 1.405 | 1.161 |
| 3 | 0.513 | 0.439 | 1.068 | 0.311 | 0.574 | 3.408 | 0.810 | 1.577 |
| 4 | 0.491 | 0.549 | 1.351 | 0.306 | 0.383 | 3.037 | 1.090 | 0.870 |
| 5 | 0.966 | 0.470 | 1.633 | 0.010 | 0.119 | 3.239 | 0.892 | 0.860 |
| 6 | 1.247 | 0.625 | 0.585 | 0.501 | 0.042 | 3.562 | 1.295 | 0.614 |
| 7 | 0.891 | 0.433 | 1.113 | 0.398 | 0.310 | 2.482 | 1.275 | 1.132 |
| 8 | 0.428 | 0.674 | 1.269 | 0.887 | 0.059 | 2.877 | 1.241 | 1.410 |
| 9 | 1.059 | 0.564 | 0.991 | 0.639 | 0.385 | 3.636 | 1.054 | 1.075 |
| 10 | 1.303 | 0.588 | 1.259 | 0.370 | 0.191 | 3.621 | 1.198 | 0.620 |

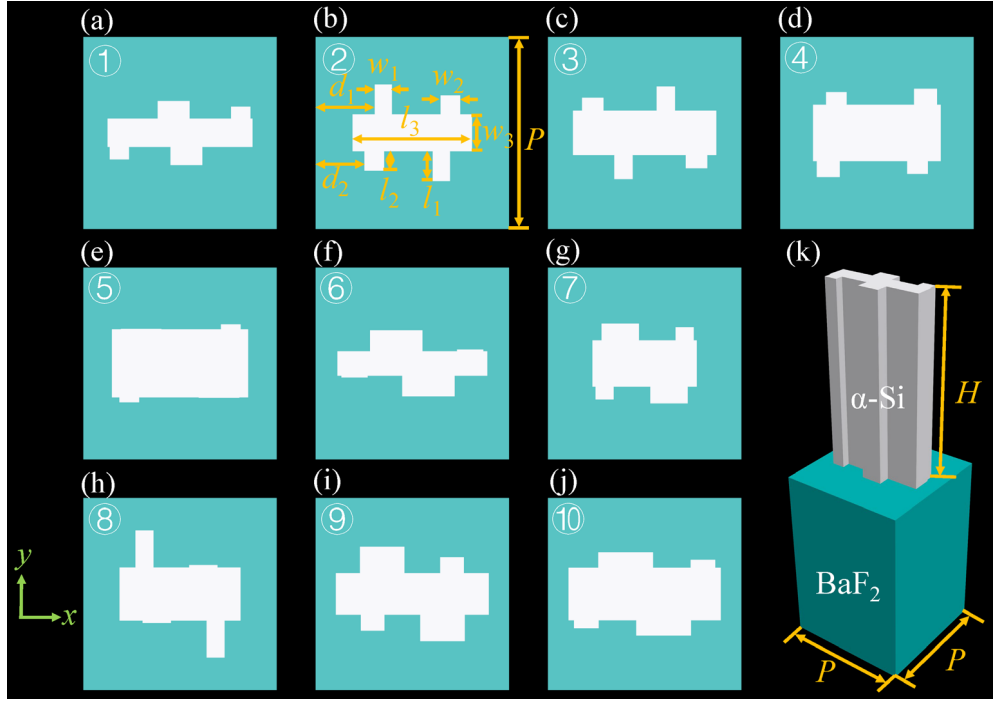

FIG. S5. (a)-(j) The top view of 10 optimized H-like shaped meta-atoms. (k) The side of the optimized meta-atom 1. The geometric parameters are marked in yellow in (b) and (k). $H$=6.8 μm, $P$=4.6 μm.

TABLE S2. The phase, polarization conversion efficiency (PCE), cross-polarization ratio (CPR) and transmission of 10 optimized meta-atoms.

|  |  | No. | 1 | 2 | 3 | 4 | 5 |
|---|---|---|---|---|---|---|---|
| RCP<br>↓<br>LCP | | Target phase (°) | 0 | 36 | 72 | 108 | 144 |
| | | Actual phase (°) | 3.73 | 30.73 | 72.20 | 110.92 | 148.68 |
| | | PCE (%) | 93.42 | 97.20 | 88.50 | 90.52 | 88.97 |
| | | CPR (%) | 99.84 | 98.79 | 92.68 | 99.02 | 99.64 |
| | | Transmission (%) | 93.57 | 98.40 | 95.49 | 91.42 | 89.29 |
| LCP<br>↓<br>RCP | | Target phase (°) | 0 | 36 | 72 | 108 | 144 |
| | | Actual phase (°) | 357.85 | 33.66 | 73.39 | 107.76 | 145.24 |
| | | PCE (%) | 94.02 | 97.52 | 88.93 | 90.98 | 89.00 |
| | | CPR (%) | 99.85 | 98.75 | 92.92 | 99.07 | 99.64 |
| | | Transmission (%) | 94.16 | 98.76 | 95.71 | 91.84 | 89.33 |
|  |  | No. | 6 | 7 | 8 | 9 | 10 |
| RCP<br>↓<br>LCP | | Target phase (°) | 0 | 36 | 72 | 108 | 144 |
| | | Actual phase (°) | 352.65 | 32.81 | 74.59 | 109.57 | 144.74 |
| | | PCE (%) | 93.08 | 92.87 | 92.29 | 87.27 | 85.57 |
| | | CPR (%) | 99.80 | 97.90 | 99.33 | 97.12 | 98.33 |
| | | Transmission (%) | 93.26 | 94.86 | 92.91 | 89.85 | 87.02 |
| LCP<br>↓<br>RCP | | Target phase (°) | 36 | 72 | 108 | 144 | 180 |
| | | Actual phase (°) | 32.66 | 70.73 | 110.02 | 147.62 | 174.94 |
| | | PCE (%) | 95.06 | 94.12 | 92.16 | 87.26 | 85.64 |
| | | CPR (%) | 99.82 | 97.78 | 99.30 | 97.08 | 98.37 |
| | | Transmission (%) | 95.22 | 96.26 | 92.81 | 89.89 | 87.06 |

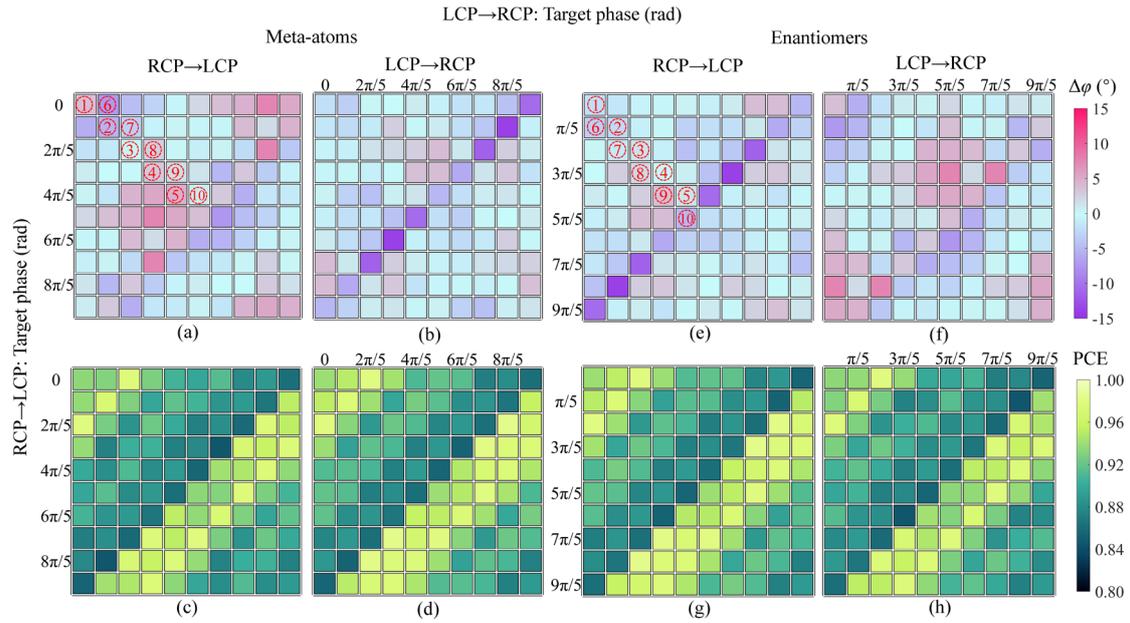

FIG. S6. (a)(b)/(e)(f) The phase difference ($\Delta\varphi$) between reality and target in the phase library (the combination of meta-atoms/enantiomers and PB phase modulation). The red number indicates the corresponding meta-atoms/enantiomers without rotation. (c)(d)/(g)(h) The map of the polarization conversion efficiency (PCE) for the meta-atoms/enantiomers.

## S7. Analysis of phase delay offset

Fig. S7(a) displays the real retardation $\delta$ of 10 kinds of H-like shaped meta-atoms, which deviate from the ideal retardation ($\delta=\pi$), resulting in a phase delay offset in comparison to the ideal PB phase delay. In Figs. S7(b) and (c), a phase delay offset of up to 6 degrees is observed when the real retardation is equal to $0.8\pi$ at a rotated angle of 45°.

Furthermore, while the phase delay of guided mode propagation is independent of the rotation angle, the arrangement within the array does impact the phase delay and PCE. Fig. S7(d) illustrates that when the meta-atom is rotated by an angle $\theta$, the arrangement is modified, resulting in a change in the phase response of guided mode propagation.

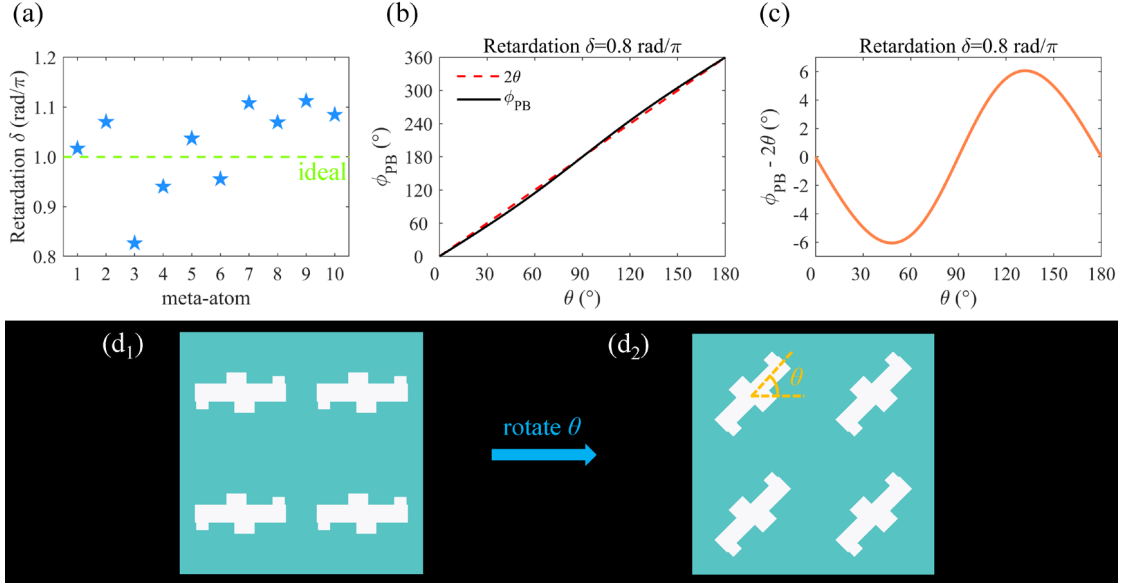

FIG. S7. (a) The retardations of 10 H-like shaped meta-atoms are marked by blue pentagrams, while the ideal retardation at $\delta=\pi$ is indicated by the green dash line. (b) The relationship between the actual $\phi_{PB}$ at a retardation of $\delta=0.8$ rad/$\pi$ (the ideal PB phase $\phi_{PB}=2\theta$ at a retardation of $\delta=1$ rad/$\pi$) and the rotation angle $\theta$. (c) The discrepancy between the actual and ideal PB phase for various rotation angles. (d) The schematic diagram of the arrangement alteration upon the rotation of the meta-atoms.

## S8. Analysis of asymmetry focus spot with linear polarization incident

For linearly polarized light incident at a polarization angle $\gamma$ with respect to the $x$-axis, the excited dipole moment in the anisotropic crystal can be expressed as:

$$\begin{bmatrix} p_x \\ p_y \end{bmatrix} = \begin{bmatrix} P_x \\ P_y \end{bmatrix} = \varepsilon_0 \begin{bmatrix} \chi_{xx} & \chi_{xy} \\ \chi_{yx} & \chi_{yy} \end{bmatrix} \begin{bmatrix} E_{xi} \\ E_{yi} \end{bmatrix} = \varepsilon_0 \begin{bmatrix} \chi_{xx} & \chi_{xy} \\ \chi_{yx} & \chi_{yy} \end{bmatrix} \begin{bmatrix} \cos\gamma \\ \sin\gamma \end{bmatrix}, \quad (39)$$

where $\vec{p} = p_x \vec{e}_x + p_y \vec{e}_y$. The transmitted electric field $\vec{E}'_t$ can be expressed as:

$$\begin{aligned} \vec{E}'_t &= A_0 \left[ p_x \vec{e}_x + p_y \vec{e}_y - \frac{1}{\sqrt{2}} \sin\xi (p_x + p_y) \vec{e}_{R_0} \right] \\ &= A_0 \varepsilon_0 [(\chi_{xx} \cos\gamma + \chi_{xy} \sin\gamma) \vec{e}_x + (\chi_{yx} \cos\gamma + \chi_{yy} \sin\gamma) \vec{e}_y \\ &\quad - \frac{1}{\sqrt{2}} \sin\xi (\chi_{xx} \cos\gamma + \chi_{xy} \sin\gamma + \chi_{yx} \cos\gamma + \chi_{yy} \sin\gamma) \vec{e}_{R_0}] \end{aligned} \quad (40)$$

The equation above indicates that the excited dipole moment in the anisotropic crystal is dependent on the incident polarization direction $\gamma$. Consequently, the radiation pattern is influenced by the incident polarization direction $\gamma$. A larger angle $\xi$ between the observation direction $\vec{e}_{R_0}$ and $z=0$ (corresponding to a larger $\sin\xi$) implies a more non-uniform radiation pattern. In summary, when linearly polarized light is incident, a larger NA of the metalens leads to larger dipolar radiation angles at the metalens edge, resulting in a more non-uniform focal point pattern.

## S9. Focus of polarization-insensitive metalenses

TABLE S3. The detailed results for the polarization-insensitive metalens with radius $R=161$ μm and focal length $f=120/370$ μm. FWHM: full width at half maximum.

| NA | 0.80 | | | |
|---|---|---|---|---|
| Incident polarization | RCP | LCP | XP | YP |
| Focal length (μm) | 118.67 | 118.67 | 119.10 | 118.46 |
| $x$-FWHM (μm) | 6.54 | 6.54 | 7.61 | 5.74 |
| $y$-FWHM (μm) | 6.54 | 6.54 | 5.74 | 7.61 |
| Focusing efficiency (%) | 76.12 | 76.64 | 77.54 | 78.13 |
| NA | 0.40 | | | |
| Incident polarization | RCP | LCP | XP | YP |
| Focal length (μm) | 359.51 | 359.51 | 356.70 | 362.64 |
| $x$-FWHM (μm) | 13.22 | 13.22 | 13.49 | 12.95 |
| $y$-FWHM (μm) | 13.22 | 13.22 | 12.69 | 13.76 |
| Focusing efficiency (%) | 82.97 | 82.74 | 82.15 | 83.94 |

TABLE S4. Comparison of focusing efficiency for long-wave infrared monochromatic polarization-insensitive metalenses. λ is the wavelength of incident light. $P$, $H$ and $H/P$ are the period, height and aspect ratio of meta-atoms, respectively. $D$, $f$ and NA are the diameter, focal length and numerical aperture of metalenses, respectively. FE is the abbreviation of Focusing efficiency.

| Year | 2021 | 2021 | 2021 | 2022 | 2022 | 2019 | 2023 | 2023 |
|---|---|---|---|---|---|---|---|---|
| λ (μm) | 9 | 9 | 10.6 | 9.28 | 10 | 10.6 | 10.6 | 10.6 |
| Structure | Si | Si | Si | Ge | Si | Si | α-Si | α-Si |
| substrate | $MgF_2$ | $MgF_2$ | Si | $BaF_2$ | Si | Si | $BaF_2$ | $BaF_2$ |
| $P$ (μm) | 3.6 | 3.6 | 3 | 4.84 | 3 | 5.5 | 4.6 | 4.6 |
| $H$ (μm) | 4 | 4 | 25.8 | 1.425 | 8 | 6.8 | 6.8 | 6.8 |
| $H/P$ | 1.11 | 1.11 | 8.6 | 0.29 | 2.67 | 1.24 | 1.48 | 1.48 |
| $D$ (μm) | 100 | 197 | 90 | 300 | 60 | 100 | 322 | 322 |
| $f$ (μm) | 88 | 85 | 50.3 | 240 | 300 | 100 | 370 | 120 |
| NA | 0.494 | 0.757 | 0.67 | 0.53 | 0.10 | 0.45 | 0.40 | 0.80 |
| FE (%) | 47.97 | 42.28 | 71 | 44.08 | 86 | 43 | 83 | 76 |
| Layer | Double | Double | Double | Double | Single | Single | Single | Single |
| Ref. | [41] | [41] | [42] | [43] | [44] | [35] | This | This |

## S10. Focus of chiral virtual-moving metalens array

TABLE S5. Detailed results of the chiral virtual-moving metalens array with side length $D$=138 μm and focal length $f$=120 μm, yielding NA=0.63.

| Incident polarization | RCP | LCP | XP | XP | YP | YP |
|---|---|---|---|---|---|---|
| Focal point position $(x, y)$ (μm) | (0, 0) | (69, 69) | (0, 0) | (69, 69) | (0, 0) | (69, 69) |
| Focal length (μm) | 120.40 | 119.09 | 120.03 | 116.27 | 121.15 | 120.21 |
| $x$-FWHM (μm) | 9.33 | 9.33 | 9.33 | 9.67 | 9.33 | 9.33 |
| $y$-FWHM (μm) | 9.33 | 9.67 | 9.33 | 9.67 | 9.33 | 9.33 |
| Focusing efficiency (%) | 84.28 | 84.22 | 42.46 | 42.67 | 44.09 | 43.73 |